\documentclass{emulateapj}
\shorttitle{the SSA 22 LAB1}
\shortauthors{Matsuda et al.}
\slugcomment{Received 2007 May 10; Accepted 2007 June 20}

\begin{document}

\title{High-Resolution Submillimeter Imaging of the Ly$\alpha$ Blob1 in SSA 22}

\author{Y. Matsuda\altaffilmark{1}, D. Iono\altaffilmark{2},
K. Ohta\altaffilmark{1}, T. Yamada\altaffilmark{2,3}, 
R. Kawabe\altaffilmark{2}, T. Hayashino\altaffilmark{4},
A. B. Peck\altaffilmark{5}, G. R. Petitpas\altaffilmark{5}}

\altaffiltext{1}{Department of Astronomy, Kyoto University, Sakyo-ku, Kyoto 606-8502, Japan; matsdayi@kusastro.kyoto-u.ac.jp}
\altaffiltext{2}{National Astronomical Observatory of Japan, 2-21-1 Osawa, Mitaka, Tokyo 181-8588}
\altaffiltext{3}{Astronomical Institute, Tohoku University, Aramaki, Aoba, Sendai 980-8578, Japan}
\altaffiltext{4}{Research Center for Neutrino Science, Tohoku University, Aramaki, Aoba, Sendai 980-8578, Japan}
\altaffiltext{5}{Harvard-Smithsonian CfA, 60 Garden St., Cambridge, MA 02138}

\begin{abstract}
We present $\sim 2\arcsec$ resolution submillimeter observations of the 
submillimeter luminous giant Ly$\alpha$ blob (LAB1) in the SSA 22 
protocluster at redshift $z=3.1$ with the Submillimeter Array (SMA). 
Although the expected submillimeter flux density is 16 mJy at
 $880\micron$, no emission is detected with the $2\farcs4 \times 1\farcs9$ 
($18\times 14$ kpc) beam at the 3$\sigma$ level of 4.2 mJy beam$^{-1}$ 
in the SMA field of view of $35\arcsec$. This is in contrast to the 
previous lower angular resolution ($15\arcsec$) observations where a 
bright (17 mJy) unresolved submillimeter source was detected at 
$850\micron$ toward the LAB1 using the Submillimeter Common-User 
Bolometer Array on the James Clerk Maxwell Telescope. The SMA 
non-detection suggests that the spatial extent of the submillimeter 
emission of LAB1 should be larger than $4\arcsec$ ($>30$ kpc). The most 
likely interpretation of the spatially extended submillimeter emission 
is that starbursts occur throughout the large area in LAB1. Some part of 
the submillimeter emission may come from spatially extended dust expelled 
from starburst regions by galactic superwind. The spatial extent of 
the submillimeter emission of LAB1 is similar to those of high redshift 
radio galaxies rather than submillimeter galaxies.

\end{abstract}


\section{INTRODUCTION}

Ly$\alpha$ blobs (LABs) are radio-quiet, giant ($30$--$200$ kpc)
Ly$\alpha$ nebulae often discovered in overdense regions of 
star-forming galaxies at high redshifts \citep{1999AJ....118.2547K,
2000ApJ...532..170S,2001ApJ...554.1001F,2004ApJ...602..545P,
2004AJ....128..569M,2005ApJ...629..654D,2006A&A...452L..23N,
2007astro.ph..3522S}. This new population of LABs may be related to 
important physical processes in galaxy formation. However, the physical 
origin of the extended Ly$\alpha$ nebulae is mysterious. Although 
similar giant Ly$\alpha$ nebulae are often seen around high 
redshift radio galaxies (HzRGs), they are thought to be related to 
their radio jets \citep[e.g.,][]{1987ApJ...319L..39M,
1997A&A...317..358V} and are presumably not the same population as 
the radio-quiet LABs. There are at least three possibilities for the 
origin of LABs: (1) Ly$\alpha$ cooling radiation resulting from 
gravitational heating \citep{2000ApJ...537L...5H,2001ApJ...562..605F,
2004MNRAS.351...63B,2006ApJ...640..539Y,2006ApJ...649...14D,
2006ApJ...649...37D,2006A&A...452L..23N}, (2) photoionization by 
metal free massive (PopIII) stars, or obscured starburst and AGN 
\citep{2000ApJ...532..170S,2001ApJ...548L..17C,2004ApJ...606...85C,
2004ApJ...614L..85B,2005ApJ...622....7F,2006Natur.441..120J}, and 
(3) shock heating by starburst-driven galactic superwind 
\citep{2000ApJ...532L..13T,2001ApJ...562L..15T,2003ApJ...591L...9O,
2004ApJ...613L..97M,2005Natur.436..227W,2005MNRAS.363.1398G,
2006Natur.440..644M}. Recent observations have revealed that LABs 
show large ($\ge$ 500 km s$^{-1}$) Ly$\alpha$ velocity widths 
\citep{2000ApJ...532..170S,
2003ApJ...591L...9O,2004MNRAS.351...63B,2005Natur.436..227W,
2006ApJ...640L.123M} and that a significant fraction of LABs 
display high luminosity infrared dust continuum emission 
\citep{2001ApJ...548L..17C,2003ApJ...583..551S,2005ApJ...629..654D,
2005MNRAS.363.1398G,2006ApJ...637L..89C}. These properties suggest 
(2) and/or (3) as the possible origin of the LABs, and a close 
relationship to the formation of massive galaxies. 

SSA 22 LAB1 at $z=3.1$ is the best suitable target to reveal the 
origin of LABs. Its spatial extent of 
$\sim 150$ kpc and a Ly$\alpha$ luminosity of $\sim 10^{44}$ ergs s$^{-1}$ 
places this as one of the largest and brightest objects among 
the known LABs \citep{2000ApJ...532..170S}. 
In a very deep Ly$\alpha$ image taken with the Subaru Prime-Focus 
Camera (Suprime-Cam), bubble like structures are seen in LAB1, which 
may be evidence for galactic superwind \citep{2004AJ....128..569M}. 
\citet{2006Natur.440..644M} 
demonstrated in their numerical simulations that such bubbles can be 
produced by galactic superwind from a proto-galaxy.
In their model, at first, multiple gas clumps are distributed 
throughout a dark matter halo. Subsequently, starbursts occur
coincidently in these clumps through interactions or mergers, 
and drive large scale gas outflows. The system evolves into an 
elliptical galaxy at the present epoch.

SSA 22 LAB1 has been observed \citep{2001ApJ...548L..17C,
2004ApJ...606...85C} with the Submillimeter Common-User Bolometer 
Array (SCUBA) on the James Clerk Maxwell Telescope. An unresolved 
submillimeter source was detected towards LAB1 with a $850\micron$ 
flux density of $17.4\pm2.9$ mJy, which makes this one of the brightest 
submillimeter 
sources at high redshifts detected to date \citep{2005ApJ...622..772C}. 
However, the angular resolution of $15\arcsec$ of the SCUBA is 
insufficient to precisely identify the optical counterpart and to 
investigate the spatial extent of the star-formation activity in LAB1. 
Although several faint optical components were detected near the center 
of LAB1 with Hubble Space Telescope (HST) observations 
\citep{2004ApJ...606...85C}, it has not been confirmed that they are 
optical counterparts to the submillimeter source yet. A radio source 
was detected with a 1.4 GHz flux density of $41.2\pm9.3~\mu$Jy 
($4\sigma$) at one of the optical 
components, J1, which is possibly the counterpart to the submillimeter 
source \citep{2004ApJ...606...85C,2005ApJ...622..772C}. However, a very 
red $K_{S}$ band source ($R-K_{S}>4.5$) was detected at another optical 
component, J2 \citep{2000ApJ...532..170S}, which is $\sim 2\arcsec$ 
from J1. Very recently, J2 was also identified as a mid-infrared 
source with Spitzer/IRAC observations and it is also possibly the 
counterpart to the submillimeter source \citep{2006astro.ph.12272G}. It 
is also possible that the submillimeter emission of LAB1 may not be 
associated with compact source(s), but has a large spatial extent due to 
galactic superwind similar to the nearby starburst galaxy, M82, in which the 
distribution of dust is spread out beyond the central (optical) starburst 
region \citep[e.g.][]{1999A&A...353...51A}. In order to examine the spatial 
extent of star-formation activity and to seek the relation between the 
production of LABs and galactic superwinds, higher angular resolution 
and more sensitive submillimeter observations are required. 

The Submillimeter Array \citep[SMA,][]{2004ApJ...616....L1} provides
an angular resolution of $\sim 2\arcsec$, which is about eight times 
higher than that with SCUBA. It is also sensitive enough 
to detect bright submillimeter sources at high redshifts in a full 
night of observation. For these reasons, the SMA is a very powerful 
instrument to examine the morphology of submillimeter sources at high 
redshifts. For example, \citet{2006ApJ...640L...1I} have successfully 
detected submillimeter emission at the $10\sigma$ level for two 20 
mJy submillimeter galaxies (SMGs), SMM J123711+6222212 and MIPS 
J142824.0+352619, using the SMA, and they obtained firm upper limits 
of $1\farcs2$ to the source sizes. 
In this letter we report a result of $\sim 2\arcsec$ resolution 
interferometric submillimeter observation of SSA22 LAB1 with the 
SMA and discuss 
the nature of the object. We assume a flat universe of $\Omega_{\rm M} 
= 0.3$, $\Omega_{\Lambda} = 0.7$, and $H_0 = 70$ km s$^{-1}$ Mpc$^{-1}$ 
($1\farcs0$ corresponds to a physical length of 7.6 kpc at $z=3.1$).

\begin{figure}[hbt]
  \plotone{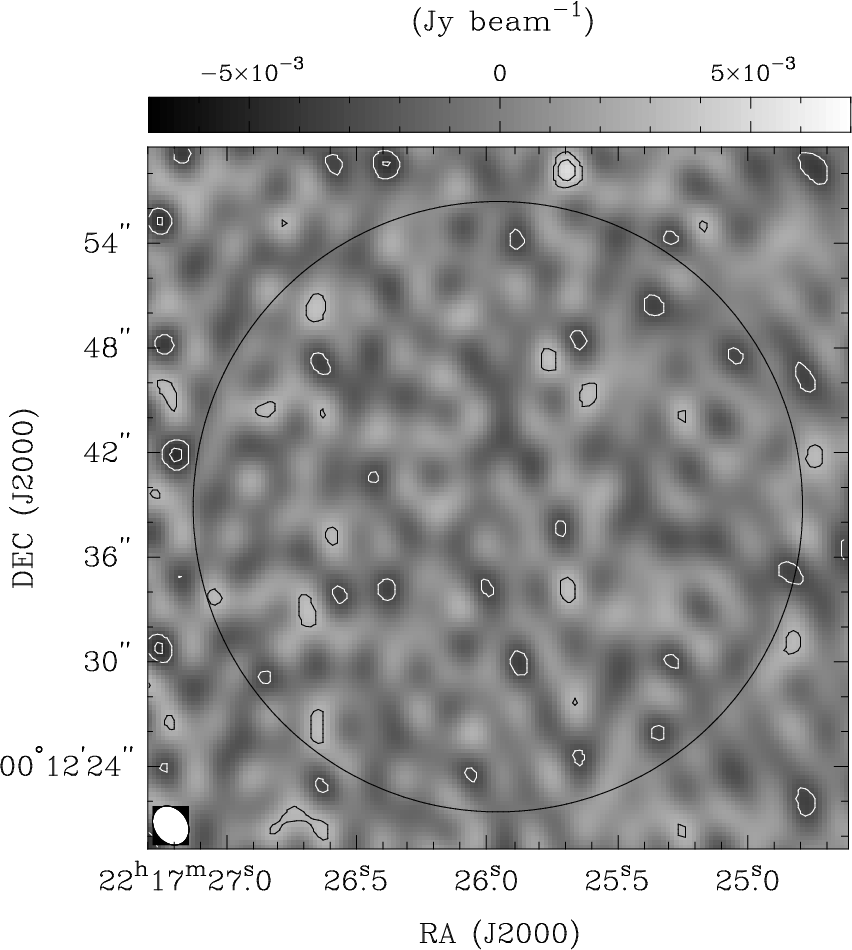}
  \caption{ Natural weighted map of the $880\micron$ continuum emission of SSA22 LAB1. The synthesized beam size has a FWHM of $2\farcs4 \times 1\farcs9$ (P.A. = $36^\circ$), which is shown in the lower left corner. The contours are at -3, -2, 2, and $3 \sigma$, with $\sigma=1.4$ mJy beam$^{-1}$. The circle shows the FWHM of the primary beam ($35\arcsec$).}  
\label{map}
\end{figure}

\section{OBSERVATIONS AND DATA REDUCTION}

Observations were carried out on 2006 September 24 (1 track, 4.5 hours 
on source) using eight antennas in the compact configuration of the
SMA\footnote{The Submillimeter Array is a joint project 
between the Smithsonian Astrophysical Observatory and the Academia 
Sinica Institute of Astronomy and Astrophysics, and is funded by the
Smithsonian Institution and the Academia Sinica.}. The SIS receivers 
were tuned to a center frequency of 345~GHz
(869$\micron$) in the upper sideband (USB), yielding 335~GHz
(895$\micron$) in the lower sideband (LSB). The SMA correlator had 
2 GHz total bandwidth in each sideband. The adopted phase reference
center of the target source was 
$\alpha$~(J2000)~$= 22^{\rm h} 17^{\rm m} 25^{\rm s}.97$, 
$\delta$~(J2000)~$=00^{\circ} 12' 38\farcs9$ \citep{2005ApJ...622..772C}. 
The track was taken under good atmospheric opacity (i.e. 
$\tau_{225} \le 0.06$). The range of (unprojected) baseline lengths 
was 16 -- 69 meters.

The SMA data were calibrated using the Caltech software package MIR, 
modified for the SMA. Antenna based passband calibration was done using 
two bright QSOs, J2225$-$0457 (3C446, 1.4~Jy) and J2148+0657 (1.0~Jy), 
as well as Uranus (69.0~Jy) and Callisto (5.0~Jy). Antenna based 
time-dependent complex gain calibration was carried out using two 
QSOs, J2225$-$0457 (1.4~Jy; 6$^\circ$ away from the target) and 
J2148+0657 (1.0~Jy; 10$^\circ$). The astrometry was checked by 
applying the phase calibration to a nearby QSO J2218$-$0335 (0.2~Jy; 
4$^\circ$) and we assess the accuracy of the astrometry to be 
$\sim 0\farcs2$. The astrometric accuracy of all of these 
calibrators is better than 10 milliarcsec, and the positional 
reference is the International Celestial Reference Frame 
\citep[ICRF,][]{1998AJ....116..516M}. Finally, absolute flux 
calibration was performed 
using Callisto. Data calibrations were carried out by two independent 
groups, which allowed us to confirm the repeatability and the robustness 
of the results. Imaging was carried out in MIRIAD \citep{miriad}. The 
natural weighted beam gave a synthesized beam size of 
$2\farcs4 \times 1\farcs9$ (FWHM, P.A. = $36^\circ$). The rms noise 
after combining the two sidebands (effectively at $882 \micron$) is 
1.4~mJy. The FWHM of the primary beam is $35\arcsec$. 

\begin{figure*}[hbt]
  \plotone{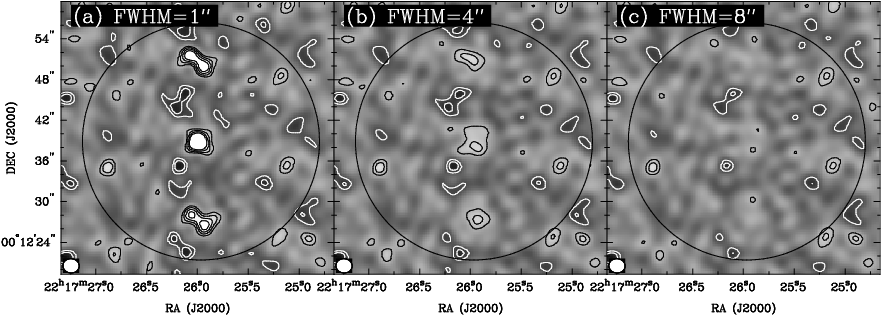}
  \caption{ Simulated SMA 880$\micron$ maps of spatially extended sources 
with a single Gaussian profile. The model sources are placed at the phase 
reference center. The FWHMs of the model sources are 1{\arcsec}, 
4{\arcsec}, and 8{\arcsec} from left to right. The contours are at -3, -2, 2, 3, 4, and $5 \sigma$, with $\sigma=1.4$ mJy beam$^{-1}$. The 
circles show the FWHM of the primary beam ($35\arcsec$). 
The components located at $\sim 10\arcsec$ north and $\sim 10\arcsec$ south 
 from the center are side-lobe response of the model source.}
  \label{sim}
\end{figure*}

\section{RESULTS}

Figure~\ref{map} shows the 880$\micron$ continuum map of SSA 22 LAB1. 
No significant emission is detected at the 3$\sigma$ limit of 4.2 mJy 
beam$^{-1}$ in the field of view of $35\arcsec$. If we adopt a dust 
temperature of $T_{d}=32.1$ K and a power-law emissivity spectral 
index of $\beta=1.5$ \citep{2005ApJ...622..772C}, the extrapolated 
$880\micron$ flux density of LAB1 is 16.2 mJy from the SCUBA
$850\micron$ flux density of 17.4 mJy. If there were a single 
unresolved source, it would be detected at $11\sigma$ level in the 
$880\micron$ map. Thus the non-detection suggests the existence of a 
spatially extended submillimeter emission associated with LAB1 assuming 
that the SCUBA detection is real. Alternatively, if there are multiple 
compact sources, the flux density 
of each source is lower than the 3$\sigma$ limit of 4.2 mJy beam$^{-1}$ 
and therefore at least four sources are needed to produce the total 
flux density of 16.2 mJy.  Widely spaced distribution of the sources 
is also required to satisfy the $3\sigma$ limit; the separation of 
the sources should be larger than the beam size of $\sim 2\arcsec$ 
(15 kpc). We note that what appears like artifacts in the map actually 
arises from high side-lobe level that extends N-S of the synthesized
beam. The synthesized beam has up to 70\% side-lobes because the source 
is at zero declination. The noise peaks in the resultant maps also 
carry this high side-lobe, resulting in structure that appears like 
diagonal ridges. The theoretical noise level derived using the 
observed SMA parameters are consistent with the observed rms of 
1.4 mJy.

In order to check the appearance of spatially extended sources in 
the SMA 880$\micron$ map, we carried out simulations of the SMA 
observations using $uvgen$ in MIRIAD (Figure~\ref{sim}). In this 
simulation, we consider three model sources, each of which has a 
single Gaussian profile with the FWHM of 1{\arcsec}, 4{\arcsec}, 
and 8{\arcsec}. We assume that the sources have a total flux density 
of 16.2 mJy intrinsically and are placed at the phase center of 
the maps. We use the same observational parameters 
(i.e. the configuration of the SMA antennas, the coordinates of the 
source, and the hour angle range) for this simulation. The rms noise of 
1.4 mJy beam$^{-1}$ in the simulated maps is the same as that of the 
observed map. The simulation shows that we could detect the source 
with FWHM of 1{\arcsec} at 10$\sigma$ level, and with FWHM of 
4{\arcsec} at 3$\sigma$ level in the observed map. However, the 
source is resolved out when the diameter of the source is larger than 
FWHM of 8{\arcsec}. Note that these simulations do not include possible 
pointing and phase calibration errors. Thus, a conservative lower limit 
of the FWHM of the spatial extent of the submillimeter emission is 
$4\arcsec$. Since the source is unresolved in the previous $15\arcsec$ 
resolution SCUBA map \citep{2001ApJ...548L..17C}, a conservative upper 
limit of the FWHM is $15\arcsec$. The FWHM of the submillimeter emission 
associated to LAB1 should range from $4\arcsec$ to $15\arcsec$, which 
corresponds to $30 - 110$ kpc at $z=3.1$, if it is indeed a single source. 

\begin{figure}
\plotone{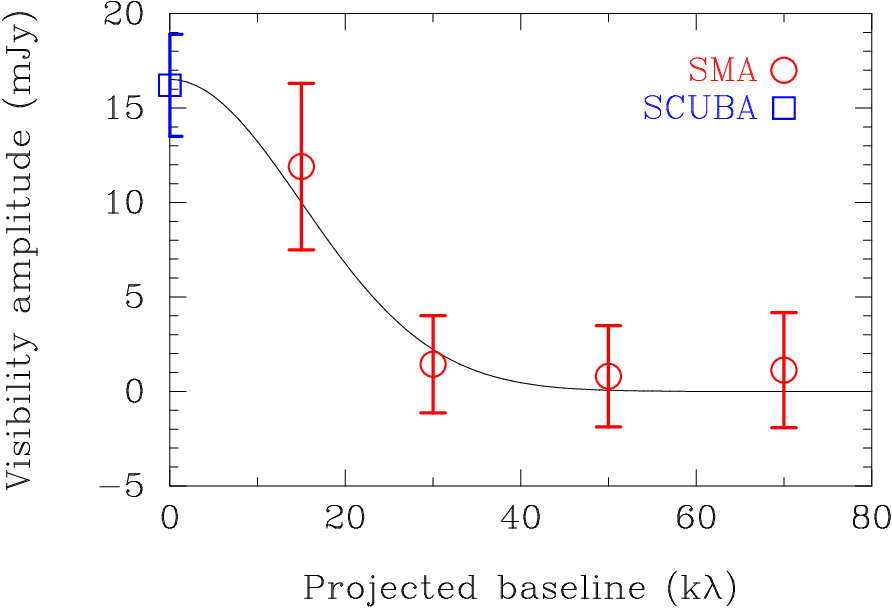}
  \caption{Visibility plot for the $880\micron$ emission. The circles show the visibility amplitude averaged in circular bins ($10-20$, $20-40$, $40-60$, and $60-80$ k$\lambda$) on $u$,$v$-plane. The vertical bars show the $1\sigma$ errors. The square shows the total $880\micron$ flux density extrapolated from the previous SCUBA observations \citep{2005ApJ...622..772C}. The solid curve is the best fit Gaussian source with a flux density of $16.5\pm2.6$ mJy and a FWHM of $5\farcs2\pm1\farcs3$.}
  \label{uv}
\end{figure}

Figure~\ref{uv} shows the visibility amplitude versus projected baseline 
with $1\sigma$ error bars. Assuming the source position is at the phase
center, we calculate the vector averaged amplitudes from 
the visibility data in circular bins of $10-20$, $20-40$, $40-60$, and 
$60-80$ k$\lambda$. Each bin contains about 3000 -- 8000 raw visibility 
points. The visibility amplitude shows a marginal (2.7$\sigma$) excess
at the bin that represents data for $10-20$ k$\lambda$. We also plot the 
total $880\micron$ flux density 
extrapolated from the SCUBA measurement ($16.2\pm2.7$~mJy) at the 
projected baseline of zero. In order to constrain the spatial extent 
of the submillimeter emission, we fit a Gaussian profile to the visibility 
data and the total flux density data. The best fit Gaussian source has a 
flux density of $16.5\pm 2.6$ mJy and a FWHM of $5\farcs2\pm1\farcs3$, 
which corresponds to $40\pm10$ kpc at $z=3.1$. The suggested source size 
of $5\farcs2$ from the visibility data is consistent with the range of
the source size of $\sim 4 - 15 \arcsec$ expected from the simulations.

\begin{figure}
\plotone{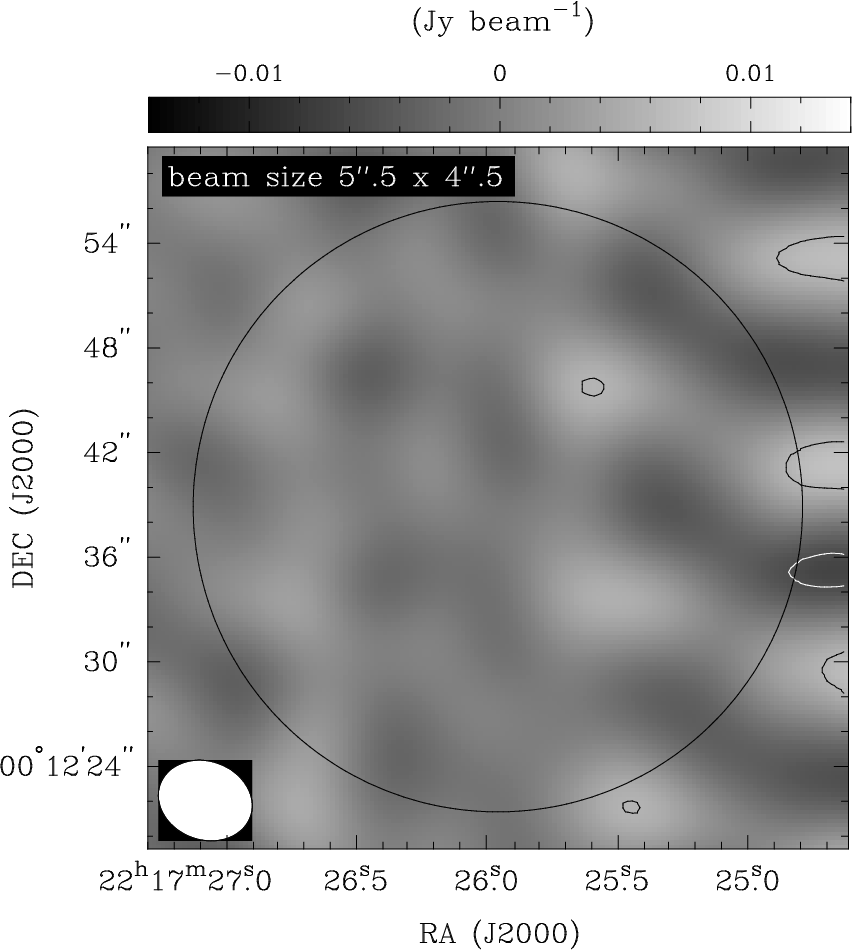}
  \caption{ Lower resolution (Gaussian tapered) $880\micron$ map of SSA22 LAB1. The synthesized beam size has a FWHM of $5\farcs5 \times 4\farcs5$ (P.A. = $70^\circ$). The contours are at -3, -2, 2, and $3 \sigma$, with $\sigma=2.8$ mJy beam$^{-1}$. The circle shows the FWHM of the primary beam ($35\arcsec$).}
  \label{taper}
\end{figure}

To increase the detectability of the extended submillimeter emission, 
we tapered the visibility data with a Gaussian that has a FWHM of 
$6\arcsec$ (Figure~\ref{taper}). The synthesized beam size of the map 
is $5\farcs5 \times 4\farcs5$ and the rms noise is 2.8 mJy beam$^{-1}$. 
However, we could not detect the extended submillimeter emission in the 
map at the 3 $\sigma$ limit of 8.4 mJy beam$^{-1}$. The peak flux 
density of 6.0 mJy beam$^{-1}$ (2.1 $\sigma$) is expected for the 
source with the FWHM of $5\farcs2$ and the total flux density of 
16.2mJy. Thus the non-detection in the lower resolution map is 
consistent with the expected peak flux density of the extended 
submillimeter emission.

In summary, the data suggest that the spatial extent of the
submillimeter emission of LAB1 should be larger than $4\arcsec$ ($>30$
kpc), if the source has a Gaussian profile. However, the S/N of the data
is too low to distinguish whether the submillimeter emission comes from
a single extended source, widely spaced multiple compact sources, or a
combination of both.

\section{DISCUSSION}

\subsection{Origin of the Submillimeter Emission}

\subsubsection{Extended Starbursts}

\begin{figure*}[hbt]
  \plotone{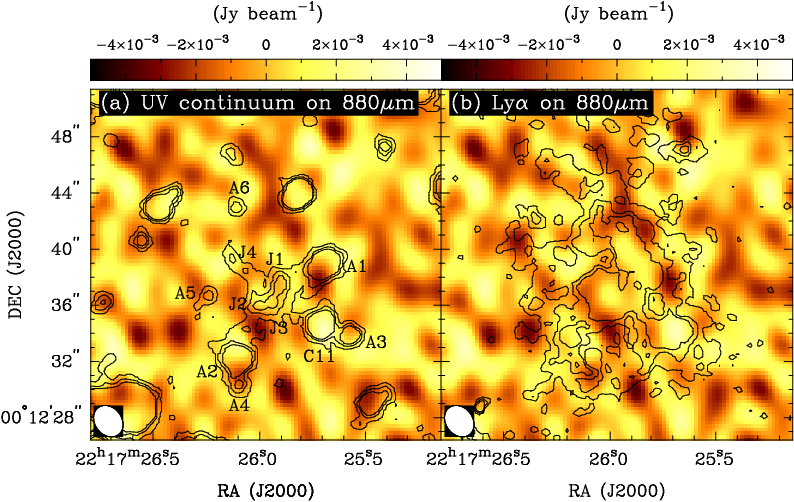}
  \caption{  ({\it a}) Subaru/Suprime-Cam $R$-band image contours
 superposed on the same $880\micron$ map. Possible optical counterparts (C11, J1 -- 4 , and A1 -- 6) are labeled. ({\it b}) Subaru/Suprime-Cam Ly$\alpha$ image contours superposed on the $880\micron$ map. The contours show 3, 6, and $9 \sigma$ per arcsec$^2$ for each image. The FWHM of the beam is shown at the bottom left. }
  \label{spcam}
\end{figure*}

We discuss possible interpretations of the spatially extended 
submillimeter emission of LAB1. One possible (and probably most 
likely) interpretation is spatially extended star-formation 
in LAB1. There is evidence to support this interpretation. It 
seems that the UV continuum emission is also spatially extended 
in LAB1. Figure~\ref{spcam}~(left) shows $R$-band (UV continuum at 
the rest frame of LAB1) image contours superposed on the 880$\micron$ 
map. The resolution of the $R$-band image is $1\farcs0$. The 
astrometric system of the image is defined by the reference 
frame of the 2MASS All-Sky Catalog of Point Sources 
\citep{2003tmc..book.....C}, and the typical rms error is 
less than $0.2\arcsec$. In Figure~\ref{spcam}~(left), we label 
UV continuum sources that are possibly associated with LAB1, 
C11 \citep{2000ApJ...532..170S}, J1 -- 4 sources 
\citep{2004ApJ...606...85C}, and A1 -- 6. The UV continuum sources 
are widely spaced and have the median separation of $\sim 6\arcsec$. 
The spectroscopic redshift of C11 was measured to be the same as LAB1 
\citep{2003ApJ...592..728S}. There is a marginal ($2.5\sigma$) excess 
of submillimeter emission at the position of C11, suggesting that C11 
is one of the counterparts of the submillimeter emission. A2 shows a 
marginal narrow-band deficit ($\sim 0.4$ mag), which suggests it has 
Ly$\alpha$ absorption at $z=3.1$ and is associated with LAB1. We derived
the photometric redshifts of the UV continuum sources inside the 
Ly$\alpha$ nebula using Palomar $U_n$-band image 
\citep{2003ApJ...592..728S} and Suprime-Cam $B$, $V$, $R$, $i'$, 
and, $z'$ bands images \citep{2004AJ....128.2073H} with Hyperz 
\citep{2000A&A...363..476B}. We found that the sources brighter than
$R=26$ magnitude (i.e. C11, J1 -- 3, and A1 -- 3) show photometric 
redshifts between $z=2.6$ and $z=3.4$. These redshifts are consistent 
with $z=3.1$, if we consider the estimated uncertainty of the 
photometric redshifts of $\Delta z\sim 0.3$. Moreover diffuse UV 
continuum emission seems to connect these sources with each other. 
The diffuse UV continuum emission may be tidal tails and bridges 
due to interactions of these sources. The photometric redshifts 
of the UV continuum sources and the connecting diffuse UV continuum 
emission suggest that most of the UV continuum sources are associated 
with LAB1. It is possible that the distribution of the submillimeter 
emission is similar to that of the UV continuum emission.

We derive a star-formation rate (SFR) in LAB1 of 1400 $M_{\odot}$ 
yr$^{-1}$ from the far-infrared (FIR) luminosity of 
$L_{\rm FIR} = 7.9 \times 10^{12} L_{\odot}$ \citep{2005ApJ...622..772C} 
and the relation between far-infrared (FIR) luminosity and SFR, 
$SFR(M_{\odot} {\rm yr}^{-1})=1.7 \times 10^{-10} L_{\rm FIR}(L_{\odot})$ 
\citep{1998ARA&A..36..189K}. The large SFR of LAB1 suggests that intense 
starbursts occur in LAB1. We also derive an average SFR surface density 
of $1.1$ $M_{\odot}$ yr$^{-1}$ kpc$^{-2}$ from dividing the SFR by the
area of the submillimeter emission, $\pi r_{\rm submm}^2$, where 
$r_{\rm submm}$ is a half size of the FWHM of the submillimeter 
emission of 40 kpc. The average SFR surface density of LAB1 is 
comparable to those of local starburst galaxies estimated from 
FIR luminosity and H$\alpha$ size \citep{1997AJ....114...54M}. 
Thus the large spatial extent of the submillimeter emission and 
the high average SFR surface density suggest that 
starbursts occur throughout the large area in LAB1.

We also derive a SFR in LAB1 of 220 $M_{\odot}$ yr$^{-1}$ from the 
UV luminosity density at $1600$ \AA\ of $L_{\nu} = 1.6 \times 
10^{30}~{\rm ergs}~{\rm s}^{-1}~{\rm Hz}^{-1}$ for the UV continuum 
emission inside the Ly$\alpha$ nebula and the relation between UV 
luminosity density and SFR, $SFR(M_{\odot}~{\rm yr}^{-1}) = 1.4 
\times 10^{-28}~L_{\nu}({\rm ergs}~{\rm s}^{-1}~{\rm Hz}^{-1})$ 
\citep{1998ARA&A..36..189K}. We do not correct for dust extinction 
in this calculation. The SFR derived from the UV continuum emission 
is about an order of magnitude lower than that derived from the 
submillimeter emission. The difference suggests that most of the 
UV continuum emission produced from star-formation activity in LAB1 
is attenuated by dust. There may be buried, unidentified UV 
continuum sources in LAB1.

It is worth noting here that recent Spitzer/MIPS $24\micron$
observations have also given us a hint that LABs have multiple dusty 
sources. \citet{2006ApJ...637L..89C} detected two or three $24\micron$ 
sources at the positions of two LABs at $z=2.4$, which are possibly 
associated with the LABs. The separations of these $24\micron$ sources 
are $60-70$ kpc, similar to or somewhat larger than the separations 
of the possible optical counterparts (i.e. UV continuum sources) of 
LAB1. The widely spaced dusty sources may be a common feature of LABs.

\subsubsection{Galactic Superwind}

Some part of the submillimeter emission can be interpreted as thermal 
emission from spatially extended dust expelled from starburst regions 
by galactic superwind. Since LAB1 shows possible 
evidence for galactic superwind (i.e bubble like structures seen in 
its Ly$\alpha$ nebula), it may have a large amount of dust entrained 
in gas outflows. Figure~\ref{spcam}~(right) shows Ly$\alpha$ 
image contours superposed on the 880$\micron$ map. While the S/N 
of the submillimeter emission is low and may not be adequate for 
conclusive determinations, the submillimeter features seem to 
correlate with the high surface brightness regions 
of Ly$\alpha$ emission. The possible correlation in distributions 
between submillimeter and Ly$\alpha$ emissions is consistent with 
this interpretation. \citet{2004MNRAS.351...63B} found a Ly$\alpha$ 
velocity gradient around C11 and argued that a bipolar outflow 
seen in M82 is a local analog for C11. In fact, M82 shows not only 
an extended H$\alpha$ emission-line nebula associated with the 
bipolar gas outflow \citep{1990ApJS...74..833H,2002PASJ...54..891O} 
but also extended dust emission \citep{1999A&A...353...51A,
2006ApJ...642L.127E}. \citet{2006ApJ...642L.127E} found that 
$8\micron$ dust emission extends to 6 kpc from the central starburst 
region of M82, and the extended emission accounts for about one-third of 
the total $8\micron$ flux with IRAC observations. Thus, it is possible 
that the thermal emission from the spatially extended dust expelled 
from starburst regions also contributes the spatially extended 
submillimeter emission of LAB1. In this picture, some of the 
extended Ly$\alpha$ and UV continuum emission of LAB1 may be 
light scattered by spatially extended dust.

\subsubsection{Other Possible Interpretations}

The Sunyaev-Zeldovich (S-Z) increment is another possible 
interpretation for the spatially extended submillimeter emission. 
There is evidence that LAB1 lies in a deep dark matter potential 
well made by the proto-cluster at $z=3.1$ 
\citep{2000ApJ...532..170S,2005ApJ...634L.125M}. The dynamical mass of 
LAB1 is estimated to be $2\times 10^{13}$ M$_{\odot}$ from the size and 
the velocity width of Ly$\alpha$ emission \citep{2004MNRAS.351...63B}. 
However, a 450$\micron$ emission with a flux density of $76\pm24$ 
mJy was detected toward LAB1 and the measured 450/850 $\micron$ 
ratio is an order of magnitude higher than the expected one from 
the S-Z increment \citep{2001ApJ...548L..17C}. Thus, the S-Z 
increment is not a likely interpretation.

The {\it non-detection} of submillimeter emission in the SMA map may be 
due to time-variability of dust emission around a buried AGN, but past 
observational results provide evidence that this is unlikely. The
spectral energy distribution of LAB1 \citep[Fig.7,][]
{2004ApJ...606...85C} suggests that the contribution of non-thermal
emission to submillimeter wavelength must be negligible. The SMA
observations of LAB1 are about five years after the SCUBA
observations. If the source shows
time-variability, the submillimeter flux density had to decrease by 
a factor of four or more in five years (one year in rest frame). 
In order to produce the time-variability in such time-scale, the 
required size of dust distribution around a nucleus is less than about 
one light year ($r \sim 0.3$ pc). Taking the FIR luminosity of 
$L_{\rm FIR} = 7.9 \times 10^{12} L_{\odot}$ and the dust temperature 
of $T_{d}=32.1$ K, we estimate the minimum source size of 
$r_{\rm min} = (L_{\rm FIR} / 4 \pi \sigma T_{d}^4)^{1/2} \sim 2$ kpc. 
For the optically thin case, the source size is estimated to be much 
larger. Thus, the minimum source size is inconsistent with the required 
source size for time-variability. 

\subsection{Comparison with Other Submillimeter Luminous Populations at High-z}

In order to investigate the relation of LAB1 to other luminous 
submillimeter populations at high redshifts such as SMGs and HzRGs, 
we compare the spatial extents of the submillimeter emission of 
these objects. 

The spatial extents of the submillimeter emission of SMGs are  
5 -- 20 times smaller than that of LAB1. \citet{2006ApJ...640..228T} 
estimated the median submillimeter source size to be 
$\le 0.5\arcsec$ (4 kpc) from $\sim 0\farcs6$ resolution millimeter 
continuum and CO imaging of eight SMGs at $z\sim 2 - 3.4$ with 
the IRAM Plateau de Bure Interferometer (PdBI). 
\citet{2006ApJ...640L...1I} directly constrain the upper limits of 
the submillimeter source size of $1\farcs2$ from $\sim 2\arcsec$ 
resolution submillimeter imaging of two luminous SMGs with the SMA. 
Although some of the SMGs are resolved into two compact sources 
\citep{2006ApJ...640..228T}, most SMGs are single compact sources. 
Thus the majority of the SMGs appear to be much more compact in 
submillimeter than LAB1.

The spatial extent of the submillimeter emission of HzRGs is similar to 
or somewhat larger than that of LAB1. High resolution millimeter observations 
with the IRAM PdBI revealed the spatially extended dust and gas in 
HzRGs. \citet{2000ApJ...528..626P} found that the millimeter and CO 
emissions in a radio galaxy 4C 60.07 at $z=3.8$ extends over $\sim 30$ 
kpc. \citet{2003A&A...401..911D} also found that evidence for spatially
resolved dust emission at scales of $0\farcs5$ to $6\arcsec$ in B3 
J2330+3927 at $z=3.1$. \citet{2003Natur.425..264S} found that five of 
seven HzRGs are resolved even with the $15\arcsec$ SCUBA beam and the 
FWHM of the submillimeter emission ranges from 50 to 250 kpc. Since 
HzRGs are known to lie in overdense regions \citep[e.g.][]
{2007A&A...461..823V} as well as LAB1, both of HzRGs and LAB1 may have
spatially extended starbursts induced by interactions and mergers of gas
clumps in overdense environments at high redshifts.

\section{CONCLUSIONS}

The $\sim 2\arcsec$ resolution submillimeter observations of SSA22 
LAB1 provide evidence that the associated submillimeter emission 
extends larger than $4\arcsec$ ($30$ kpc). The most likely
interpretation for the extended submillimeter emission is spatially
extended star-formation 
in LAB1. The derived SFR surface density suggests that starbursts 
occur throughout the large area in LAB1. The hint of a possible 
correlation in distributions between submillimeter and Ly$\alpha$ 
emissions suggests that the thermal emission from the spatially 
extended dust expelled from starburst regions by galactic superwind 
may contribute the spatially extended submillimeter emission of LAB1. 
The spatial extent of submillimeter emission of LAB1 is similar to 
those of HzRGs rather than SMGs. Further data will be required 
to determine whether spatially extended submillimeter emission is 
a common feature of LABs and to investigate the relationship 
between radio-quiet LABs and HzRGs.

\acknowledgments
We thank Takashi Hattori, Tomoki Saito, Masao Mori, Keiichi Umetsu,
Koichiro Nakanishi, Masakazu Kobayashi, Thomas Greve, Masayuki Umemura,
and Hiroyuki Hirashita for insightful discussion. We also thank Ian
Smail for encouragement. This work is supported
by the Grant-in-Aid for the 21st Century COE "Center for Diversity and
Universality in Physics" from the Ministry of Education, Culture,
Sports, Science and Technology (MEXT) of Japan.

\end{document}